\documentclass[reprint,prx,aps,amsmath,amssymb,floatfix,superscriptaddress,longbibliography]{revtex4-2}

\usepackage{mathtools,amsmath}
\usepackage{graphicx} 
\usepackage[breaklinks=true,colorlinks,citecolor=teal,linkcolor=teal,urlcolor=teal]{hyperref}
\usepackage{physics}
\usepackage{siunitx}
\usepackage{bbold}
\usepackage{soul}
\usepackage{bm}
\usepackage[normalem]{ulem}
\usepackage{times}
\usepackage{enumerate}  
\usepackage{float}
\usepackage{amssymb}
\usepackage{xcolor} 
\usepackage{subfigure}

\begin{document}

\title{Quasi-degenerate resonant eigenstate doublets of two quantum emitters in a closed waveguide}

\author{Ammara Ammara}
\affiliation{Dipartimento Interuniversitario di Fisica, Universit\`a di Bari, I-70126 Bari, Italy}
\affiliation{INFN, Sezione di Bari, I-70125, Bari, Italy}

\author{Paolo Facchi}
\affiliation{Dipartimento Interuniversitario di Fisica, Universit\`a di Bari, I-70126 Bari, Italy}
\affiliation{INFN, Sezione di Bari, I-70125, Bari, Italy}

\author{Saverio Pascazio}
\affiliation{Dipartimento Interuniversitario di Fisica, Universit\`a di Bari, I-70126 Bari, Italy}
\affiliation{INFN, Sezione di Bari, I-70125, Bari, Italy}

\author{Francesco V. Pepe}\email{francesco.pepe@ba.infn.it}
\affiliation{Dipartimento Interuniversitario di Fisica, Universit\`a di Bari, I-70126 Bari, Italy}
\affiliation{INFN, Sezione di Bari, I-70125, Bari, Italy}

\author{Debmalya Das}
\affiliation{Dipartimento Interuniversitario di Fisica, Universit\`a di Bari, I-70126 Bari, Italy}
\affiliation{INFN, Sezione di Bari, I-70125, Bari, Italy}

%\date{January 2025}

\begin{abstract}
   The physics of systems of quantum emitters in waveguide quantum electrodynamics is significantly influenced by the relation between their spatial separation and the wavelength of the emitted photons. If the distance that separates a pair of emitters meets specific resonance conditions, the photon amplitudes produced from decay may destructively interfere. In an infinite-waveguide setting, this effect gives rise to bound states in the continuum, where a photon remains confined between the emitters. In the case of a finite-length waveguide with periodic boundary conditions, there exist two such relevant
distances for a given arrangement of the quantum emitters, leading to states in which a photon is confined to either the shorter or the longer path that connects the emitters. If the ratio of the shorter and the longer path is a rational number, these two kinds of resonant eigenstates are allowed to co-exist for the same Hamiltonian. In this paper, we investigate the existence of quasi-degenerate resonant doublets of a pair of identical emitters coupled to a linear waveguide mode. The states that form the doublet are searched among the ones in which a single excitation tends to remain bound to the emitters. We investigate the spectrum in a finite range around degeneracy points to check whether the doublet remains well separated from the closest eigenvalues in the spectrum. The identification of quasi-degenerate doublets opens the possibility to manipulate the emitters-waveguide system as an effectively two-level system in specific energy ranges, providing an innovative tool for quantum technology tasks.
\end{abstract}

\maketitle

\section{Introduction}

In recent times, systems of quantum emitters interacting with a linear waveguide have emerged as some of the most interesting one-dimensional platforms to study the interplay of light and matter~\cite{Roy2017, Facchi2016, Facchi2018a, Facchi2018b, Facchi2019, Zhang2019, Calajo2019, Bello2019, Maffei2024, Magnifico2025, Lonigro2021a, Lonigro2021b}. The~explored configurations include waveguides with single, two, or~multiple quantum emitters~\cite{Burillo2017, Lalumiere2013, Tufarelli2013, Witthaut2010, Tudela2011, Facchi2016, Facchi2018a, Zhang2019, Zheng2013, Ballestero2013, Redchenko2014, Arjan2013, Calajo2019, Facchi2018b, Sinha2020, Yudson2008, Fang2015, Tsoi2008, Bello2019, Maffei2024, Magnifico2025, Lonigro2021a, Lonigro2021b, Kockum2018, Facchi2019, Zhou2017}. The~ability to control multiple parameters such as emitter-field couplings, emitter-emitter couplings, inter-emitter distances, and~excitation frequencies, along with one-dimensional confinement, makes these systems an attractive prospect for theoretical explorations and investigations concerning technological implications~\cite{Tudela2011, You2011, Faraon2007, Vetsch2010, Bajcsy2009, Wallraff2004, Astafiev2010, Douglas2015, Shen2005}. The~physics of systems composed of two or more two-level emitters (qubits) is significantly influenced by the relation between the spatial separation of the emitters and the wavelength of the emitted electromagnetic wave. It is of great interest that when the distance between two emitters meets certain conditions, photons produced from decay may destructively interfere, giving rise to a bound state in the continuum (BIC) \cite{Stillinger1975, Friedrich1985}. In~the weak-coupling regime, the~reduced state of the qubits for a BIC is to a good approximation a Bell~state. 

The generations of Bell states and BICs are thus intimately connected. Starting from a separable state of two qubits and a field with a single total excitation, a~superposition of a Bell state with no excitation in the field and a BIC with the qubits in the ground state can be obtained via entanglement by relaxation~\cite{Facchi2016}. In~other interesting situations, the~distance between the emitters can be long enough necessitating a non-Markovian treatment~\cite{, Ballestero2013, Facchi2016, Sinha2020, Sinha2020b, Trivedi2021}, time-dependent Hamiltonians~\cite{Hayran2021} or two-photon scattering~\cite{Calajo2019, Trivedi2021}. Recently, a~Bell state and a BIC have been generated by sending one photon in a waveguide with two emitters in the ground state~\cite{Magnifico2025}. 

In the case of a closed waveguide forming a ring of length $L$, coupled to two emitters, one can find \textit{resonant} %MDPI: Please confirm if the italics are necessary; if not, please remove them. 
%AUTHORS: We confirm that it is necessary, we prefer not to remove it here.
 eigenstates of the joint emitter-photon Hamiltonian whose eigenvalues correspond, through the field dispersion relation, to~the frequency of a stationary wave confined in one of the two waveguide segments that separate the emitters~\cite{Lonigro2021a}. A~particularly interesting situation occurs when the energy of a stationary photon confined in the shorter segment between the two emitters, of~length $d$, coincides with the energy of a photon confined in the longer segment between the emitters, of~length $L-d$. In~Ref.~\cite{Lonigro2021a}, it was conjectured that a quasi-degenerate pair of eigenstates, characterized by the photon amplitude concentrated in the shorter and in the longer path, can occur in the spectrum of the system Hamiltonian. If~the emitter excitation is predominant over the field excitation, the~eigenstates related to such a pair of energies should be characterized by opposite spatial symmetry with respect to the midpoint between the emitters. Under~certain conditions, such an energy separation could be suitable to coherently address and manipulate the system as effectively two-level (logical qubit)~\cite{Lonigro2021a}. Two-level systems that are robust, yet easy to coherently manipulate using external driving fields form the backbone of quantum computing architectures~\cite{nielsenchuang,benenti}. To~fully fathom the utility of these complementary pairs of resonant states, it is therefore extremely necessary to characterize~them.

In this paper, we investigate the existence of quasi-degenerate resonant doublets in the following way. First, we consider a pair of identical emitters coupled to a waveguide in such a way that $d/(L-d)$ is a rational number. In~this case, the~shorter and longer path can be at the same time multiple of a given photon half-wavelength. Then, we explore the values of the emitter excitation energy in which a resonant doublet is expected to occur. For~each of these values, we diagonalize the Hamiltonian and identify the two eigenstates with the highest emitter excitation probability. We search for the cases in which both these two eigenstates have degenerate eigenvalues and photon amplitude profiles that correspond to the expectation for resonant eigenstates of opposite symmetry. Finally, we determine if, in~a given range of excitation energy around the found degeneracy point, the~resonant doublet actually remains quasi-degenerate, by~comparing their energy with the closest eigenvalues in the~spectrum.

The paper is organized as follows. In~Section~\ref{sec:hamiltonian}, we outline the physical features and formal aspects of the model that describes the considered system. In~Section~\ref{sec:doublets}, we apply the aforementioned workflow to identify resonant doublets for two different values of the ratio $d/(L-d)$. We finally discuss the results and their implications in Section~\ref{sec:conclusions}.

\section{Model and~Symmetries}\label{sec:hamiltonian}

We consider a closed waveguide of length \(L\),
with two quantum emitters placed at positions \(x_1\) and \(x_2\), separated by a distance $d= |x_1-x_2|$, with $0<d<L/2$. We work under the assumption that for the closed waveguide, the~only relevant factor is topology, while the specific geometry does not produce qualitatively significant effects. We describe the emitters as two-level systems, whose Hilbert space is spanned by the states $\{|g_\alpha\rangle,|e_\alpha\rangle\}$, with~$\alpha=1,2$, characterized by identical physical features. The~emitters mutually interact exclusively through the exchange of photons that belong to a single transverse mode propagating through the waveguide. Due to the periodic boundary conditions, the~allowed longitudinal components of the photon wave vector are discrete and given by $q_k=2\pi k/L$, with~$k \in \mathbb{Z}$. The~corresponding frequency values are given by
\begin{equation}
    \omega_k = \sqrt{v^2 \left(\frac{2\pi k}{L}\right)^2 + m^2} ,
\end{equation}
where $m$ is the cutoff due to the transverse confinement of the mode and $v$ is the speed of light in the unconstrained medium~\cite{jackson}. In~the following, we shall fix $\hbar=v=1$ and use $m$ as the energy unit of~measure.

In the absence of emitter-field interaction, the~system is described by the free Hamiltonian
\begin{equation}
    H_{0} = \sum_{\alpha=1,2}\varepsilon\sigma^{+}_\alpha \sigma^{-}_\alpha + \sum_{k=-\infty}^{\infty} \omega_{k} b_{k}^{\dagger} b_{k},
\end{equation}
where $\sigma^{+}_\alpha = |e_\alpha\rangle \langle g_\alpha|$ and $\sigma^{-}_\alpha = |g_\alpha\rangle \langle e_\alpha|$ are the ladder operators on the emitter $\alpha$, and~$\{b_k,b^{\dagger}_k\}$ represent the longitudinal mode operators satisfying the canonical commutation relations $[b_k, b_{k^\prime}]=0$ and $[b_k, b_{k^\prime}^\dagger]=\delta_{k k^\prime}$. The~vacuum $|\mathrm{vac}\rangle$ is defined as the state satisfying $b_k |\mathrm{vac}\rangle=0$ for all $k$'s, while single-photon states with a well-defined wavenumber are obtained as $|k\rangle = b_k^{\dagger} |\mathrm{vac}\rangle$. 

Assuming a rotating-wave approximation, the~Hamiltonian describing interaction between the emitters and the field can be modeled as
\begin{equation}
 H_{\mathrm{int}} = \sum_{k=-\infty}^{\infty} \sum_{\alpha=1,2} F_{k} \left( e^{\frac{2 \pi i k x_{\alpha}}{L}} \sigma_{\alpha}^{+} b_{k} + e^{\frac{-2 \pi i k x_{\alpha}}{L}} \sigma_{\alpha}^{-} b_{k}^{\dagger} \right) .
\end{equation}
In the %MDPI: Please check through the paper if indentation should be added to the first line after equations. Same as below.
%AUTHORS: Indentations are ok as they are in this version
following, we will use the interaction form factor
\begin{equation}
    F_k = \sqrt{\frac{\gamma}{L \omega_k}} ,
\end{equation}
depending on a constant $\gamma$, which naturally emerges in a ``$\bm{p}\cdot\bm{A}$'' %MDPI: Please confirm if the bold formatting is necessary; if not, please remove it.
%AUTHORS: The bold formatting indicates that the quantities are vectors, so it is necessary.
 dipole picture~\cite{cohentannoudji}. Such a form factor has the advantage of not requiring any additional cutoff for a one-dimensional field propagation (see Ref.~\cite{Facchi2016}).

\begin{figure}[]
%    \centering
    \includegraphics[width=0.35\textwidth]{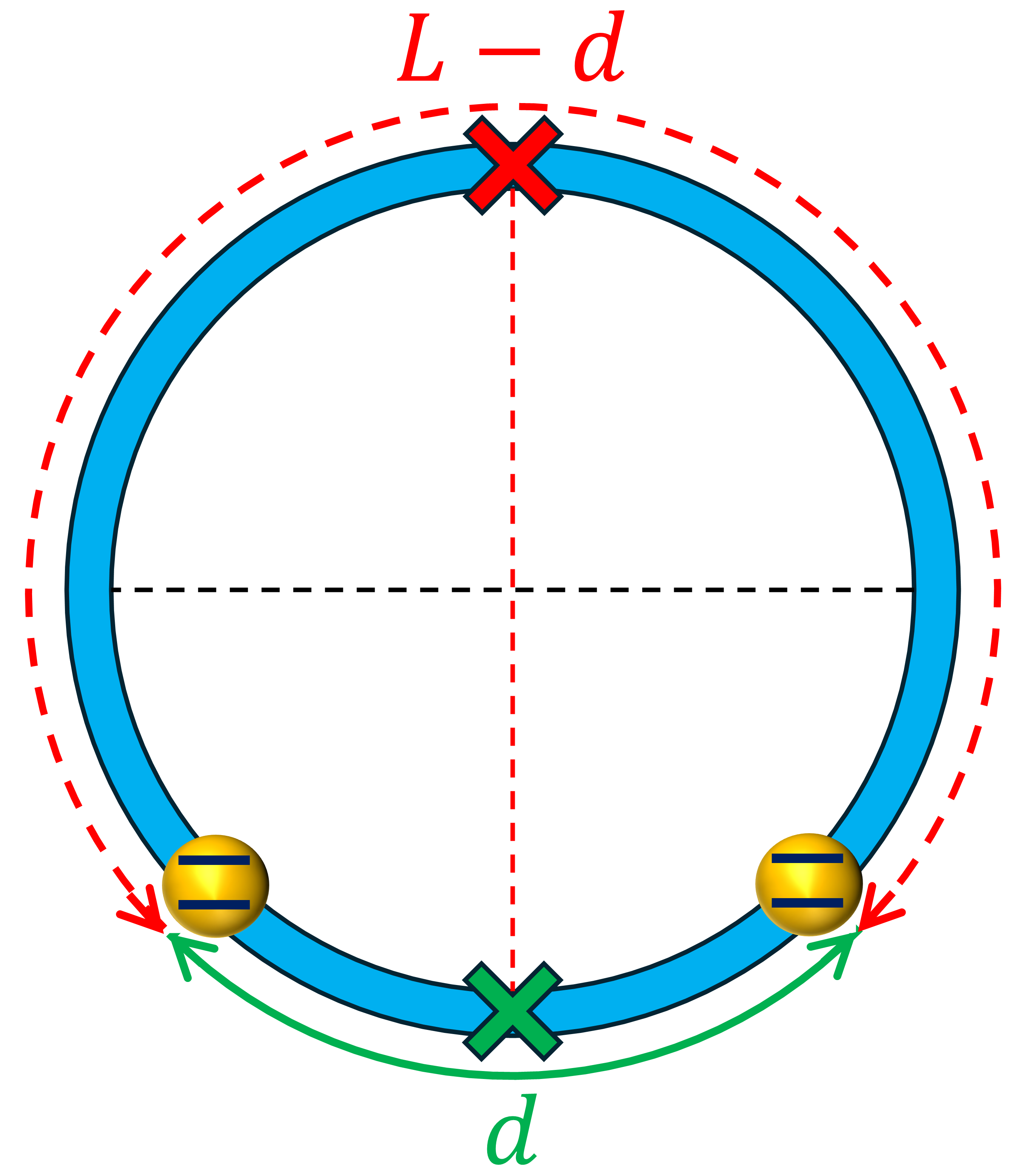}
    \caption{Schematic representation of a pair of emitters, described as two-level systems, coupled to a closed waveguide of length $L$. The~distance $d$ is the smaller between the emitters ($0<d<L/2$). The~system is symmetric around the vertical line that joins the two mid-points between the emitters (crosses). In~the special case $d=L/2$, not considered in this work, the~system acquires an additional mirror symmetry along the horizontal~line.}
    \label{fig:system}
\end{figure}

The total Hamiltonian
\begin{equation}
    H = H_0 + H_{\mathrm{int}}
\end{equation}
has relevant symmetry properties. First, it preserves the total number of excitations $N = \sum_{\alpha=1,2} \sigma_{\alpha}^+ \sigma_{\alpha}^- + \sum_{k} b_k^{\dagger}b_k$. Here, following Ref.~\cite{Lonigro2021a}, we shall focus on the $N=1$ sector, where an arbitrary state can be expanded as
\begin{equation}
    |\Psi\rangle =\sum_{\alpha=1,2} a_{\alpha} \sigma_{\alpha}^+ |G\rangle  \otimes |\mathrm{vac}\rangle + |G\rangle \otimes \sum_{k=-\infty}^{\infty} \xi_{k} b_{k}^{\dagger} |\mathrm{vac}\rangle,
\end{equation} 
where $|G\rangle=\left|g_{1}\right\rangle \otimes\left|g_{2}\right\rangle$ is the composite ground state of the free emitters, and $a_{\alpha}$ and $\xi_k$ are complex numbers that satisfy $\sum_{\alpha=1,2} |a_{\alpha}|^2 + \sum_{k=-\infty}^{\infty} |\xi_k|^2 = 1$. Moreover, as~schematically shown in Figure~\ref{fig:system}, the~system is symmetric for modular reflections with respect to the midpoint between the emitters:
\begin{equation}
    \sigma_1^{\pm} \leftrightarrow \sigma_2^{\pm}, \quad b_k \leftrightarrow b_{-k} e^{- 2\pi i k \frac{x_1+x_2}{2 L} } .
\end{equation}
This symmetry entails that the total Hamiltonian $H$ can be independently diagonalized in a symmetric sector, where the eigenstates are characterized by \begin{equation}
    a_2 = a_1 ,
\end{equation}
and an antisymmetric sector, with~\begin{equation}
    a_2 = - a_1 .
\end{equation}
The emitter excitation amplitudes of the eigenstates can be used to quantify the total emitter excitation probability
\begin{equation}\label{eq:Pe}
    P_e = |a_1|^2 + |a_2|^2 = 2 |a_1|^2 ,
\end{equation}
which represents the most relevant figure of merit for our analysis. The~spatial symmetry is reflected in the spatial amplitude of the photon~\cite{Lonigro2021a}
\begin{equation}\label{eq:amplitude}
    \zeta(x)=a_1 \xi_1(x-x_1)+a_2 \xi_1(x-x_2) ,
\end{equation}
where
\begin{equation}
    \xi_{1}(x)=\frac{\sqrt{2 \pi \gamma}}{L} \sum_{k=-\infty}^{\infty} \frac{1}{\sqrt{\omega_{k}}\left(E-\omega_{k}\right)} \exp \left(\frac{2 \pi i k x}{L}\right) ,
\end{equation}
which provides an alternative and visually convenient representation of the set $\{\xi_k\}$ of photon mode amplitudes. The~physical meaning of the spatial amplitude is related to its square modulus, proportional to the field energy density~\cite{Facchi2016}. Previous findings, obtained in both an infinite~\cite{Facchi2016} and a finite waveguide framework~\cite{Lonigro2021a}, indicate that \textit{resonant eigenstates}, in~which the electromagnetic field energy is concentrated in one of the two paths between the emitters, can occur whenever one of the distances $d$ or $L-d$ can contain an integer number of half-wavelengths corresponding to the eigenvalue. As~a consequence, to~obtain two resonant eigenstates with the same energy, the~ratio of the two inter-emitter path lengths must necessarily be rational. Remarkably, periodic boundary conditions are crucial to the availability of both paths to confine a photon between the emitters. Therefore, in~our case, we expect the eigenstates with energy equal to either
\begin{equation}\label{eq:short_E} 
    E_\nu=\sqrt{\left(\frac{\nu\pi}{d}\right)^2 + m^2} \quad \text{with } \nu \in \mathbb{N}
\end{equation}
or
\begin{equation}\label{eq:long_E}
    \tilde{E}_{\nu'}=\sqrt{\left(\frac{\nu'\pi}{L-d}\right)^2 + m^2}, \quad \text{with } \nu' \in \mathbb{N}
\end{equation}
to be resonant. At~$\gamma=0$, the~spectrum of the one-excitation Hamiltonian is made of infinite doubly degenerate eigenvalues, corresponding to photons with energy $\omega_k>m$ and spacing $\Delta\omega \simeq \sqrt{\omega_k^2 - m^2}/\omega_k$, a~nondegenerate eigenvalue $m$ corresponding to a photon state at the propagation threshold, and~a doubly-degenerate eigenvalue $\varepsilon$, corresponding to emitter excitations. In~a small-coupling picture, resonant states occur when $\varepsilon$ is close to one of the photon energies $\omega_k>m$. Therefore, resonant doublets are expected to emerge from a quasi-fourfold-degenerate sector. The~resolution of this degeneracy is generally difficult to predict, depending on specific features of the Hamiltonian such as the coupling strength, and~is crucial in determining the energy separation between the resonant doublet energies and the closest eigenvalues. We remark that, due to the symmetries, other accidental degeneracies in the coupled Hamiltonian can occur, unrelated to resonant~doublets.

It is finally worth noting that, in~the special case $d=L-d=L/2$, the~system acquires an additional symmetry (see Figure~\ref{fig:system}), associated with modular reflections with respect to any of the emitters. For~the sake of simplicity, we shall not consider this particular degenerate case in our~analysis.

\section{Doublets of Resonant~Eigenstates}\label{sec:doublets}

In view of the general properties outlined in Section~\ref{sec:hamiltonian}, we shall now investigate the existence and features of degenerate, or~quasi-degenerate, doublets of resonant eigenstates. Resonant eigenstates are expected, at~least in a small-coupling regime, to~be characterized by a predominant share of emitter excitation (namely, a~high $P_e$), due to the destructive interference of the field propagating from each emitter~\cite{Facchi2016,Calajo2019}. Therefore, the~first step in finding resonant doublets is represented by diagonalizing the Hamiltonian and identifying the two eigenstates with the highest $P_e$. Due to the constraints on the orthogonality of states belonging to the same symmetry sector, impossible if the sum of the emitter excitation probability exceeds $1$, these two high-$P_e$ eigenstates typically belong to different symmetry sectors, thus being orthogonal to each other by construction. The~second step of our search is to compare the eigenvalues corresponding to the eigenstates with the highest $P_e$ with the resonance energies \eqref{eq:short_E} and \eqref{eq:long_E}. Identification of resonant eigenstates is corroborated by studying the photon amplitude \eqref{eq:amplitude}.

For definiteness, throughout the analysis, we shall fix the coupling constant and the waveguide length to $\gamma=10^{-4}$ and $L=40\pi$, respectively (recall that the energies are expressed in units of m and lengths in units of m$^{-1}$). The emitter positions are set at $x_1=0$ and $x_2=d$. The~emitter excitation energy, instead, will be varied in the range
\begin{equation}\label{eq:range}
    \varepsilon\in [1.005,1.030],    
\end{equation}
which, as~we shall see in the following, is sufficient to identify a diversified~phenomenology.

\subsection{Case 1: $d/(L-d)=1/4$}

\begin{figure}[]
%    \centering
    \subfigure[~$\varepsilon=1.008$, $E=1.0080$, $P_e=0.8648$]{\includegraphics[width=0.22\textwidth]{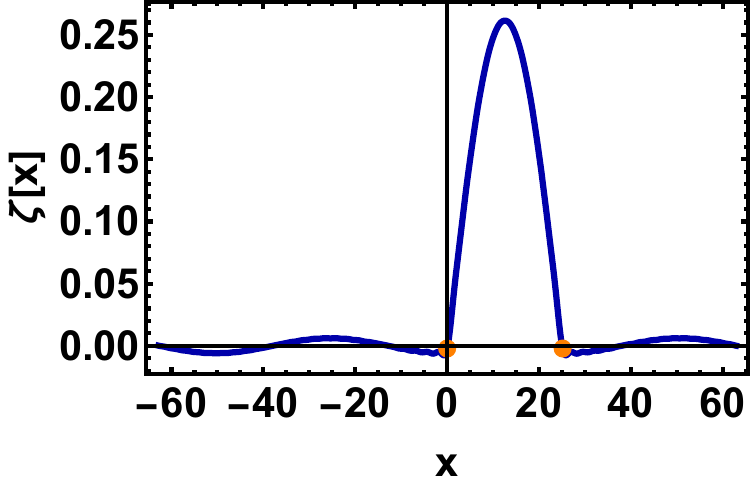}}
    \subfigure[~$\varepsilon=1.008$, $E=1.0079$, $P_e=0.6122$]{\includegraphics[width=0.22\textwidth]{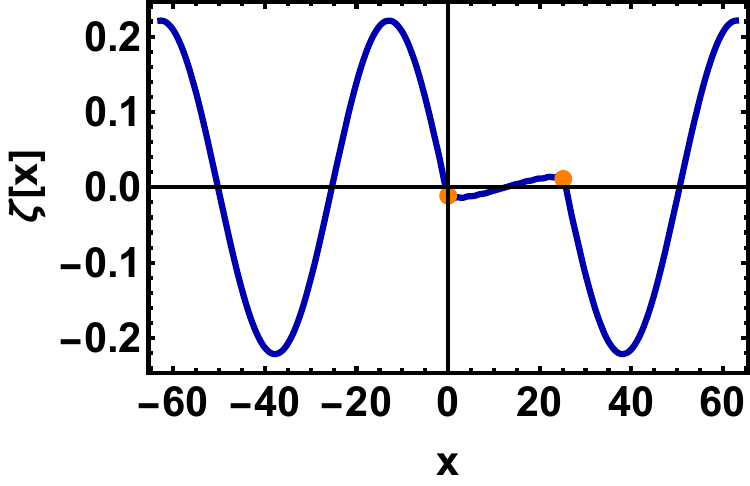}}\\
    \subfigure[~$\varepsilon=1.011$, $E=1.0105$, $P_e=0.6090$]{\includegraphics[width=0.22\textwidth]{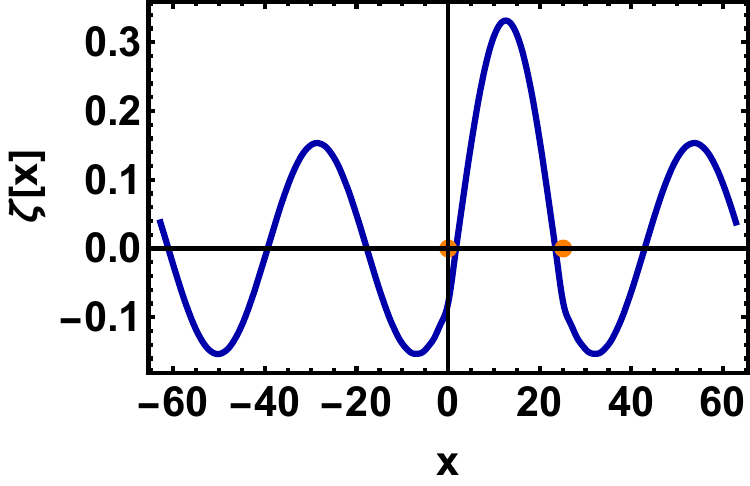}}
    \subfigure[~$\varepsilon=1.011$, $E=1.0096$, $P_e=0.4329$]{\includegraphics[width=0.22\textwidth]{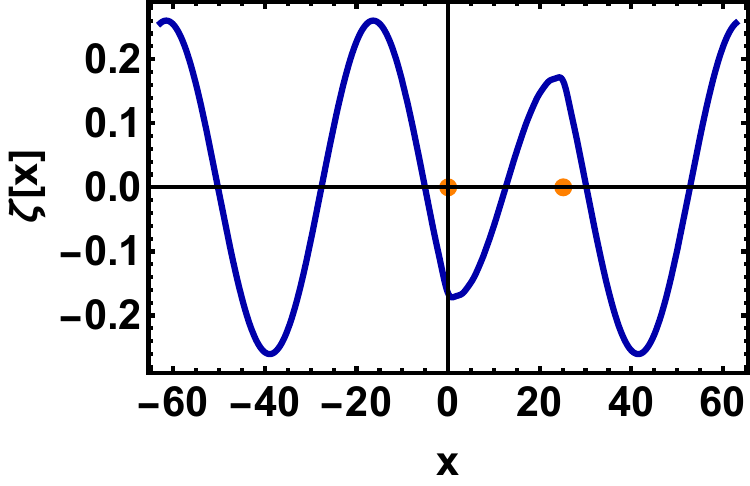}}
    \caption{Photon amplitudes corresponding to the two eigenstates with the highest emitter excitation probability for $\gamma= 10^{-4}$, $L=40\pi$ and $d=8\pi$. The~emitter excitation energy corresponds to $\varepsilon=1.008$ in the upper panels and $\varepsilon=1.011$ in the lower panels. The~eigenvalue $E$ and the emitter excitation probability $P_e$ that characterize each state are reported in the plot label. The~two states found for each $\varepsilon$ belong to sectors of opposite central symmetry with respect to the midpoint $(x_1+x_2)/2=4\pi$: this is a common tendency, but~not a general feature. In~the upper panels, the~photon amplitude (hence, the~electromagnetic energy) tends to concentrate either in the shorter [panel (\textbf{a})] or in the longer [panel (\textbf{b})] path between the emitters. In~the lower panels, there is no such concentration effect, and~the gap between the eigenvalues is~larger.}
    \label{fig:ph-wv1}
\end{figure}

As a first case study, we consider the one with
\begin{equation}
    d=8\pi, \quad L-d=32\pi ,
\end{equation}
determining a ratio of $1/4$ of the shorter and longer paths between the emitters. Let us start with an exploratory analysis, in~which we diagonalize the Hamiltonians corresponding to the two excitation energies $\varepsilon = 1.008$ and $\varepsilon = 1.011$, identify the two states characterized by the highest $P_e$ in the two cases, and~compute their related photon amplitudes. The~results, shown in Figure~\ref{fig:ph-wv1}, visually highlight a similarity between the photon amplitude profiles, related to the fact that, for~each $\varepsilon$, the~two states have opposite symmetry with respect to the midpoint $(x_1+x_2)/2 = 4\pi$. However, there is also a striking difference. In~the case $\varepsilon = 1.008$ (upper panels), the~photon amplitude is concentrated by a large amount either in the shorter or in the longer path between the emitters. In~contrast, for~$\varepsilon = 1.011$ (lower panels), there is no such confinement effect. A~subtler difference is related to the energy gap between the two states, which in the first case is smaller by an order of~magnitude with respect to the second case.

%\unskip

The observed discrepancies are related to the fact that $\varepsilon = 1.008$ is very close to a double resonance: $E_1$ on the shorter path, $\tilde{E}_4$ on the longer path. Therefore, for \mbox{$\varepsilon = 1.008$,} the~two states with the highest emitter excitation probabilities (whose energies are intuitively expected to be close to $\varepsilon$) are quasi-resonant: in case of the symmetric state, the~shorter path approximately accommodates one half-wavelength ($\nu=1$), while for the antisymmetric state almost four half-wavelengths ($\nu'=4$) fit in the longer path. In~addition, corresponding to both these states, the~photon amplitudes are very small (but not zero) in the respective complementary segments between the emitters. On~the other hand, a~similar phenomenology is not observed for the excitation energy $\varepsilon = 1.011$, which is far from any other resonance~value.

At this point, we perform a systematic analysis to determine~whether 
\begin{itemize}
    \item for some value of $\varepsilon$, the~eigenvalue of one or both states with the largest $P_e$ exactly matches a resonance energy;
    \item there exist a possible degenerate or quasi-degenerate doublet of resonant states, well-separated in energy from the rest of the spectrum.
\end{itemize}

\begin{figure}[]
%    \centering
    \includegraphics[width=0.45\textwidth]{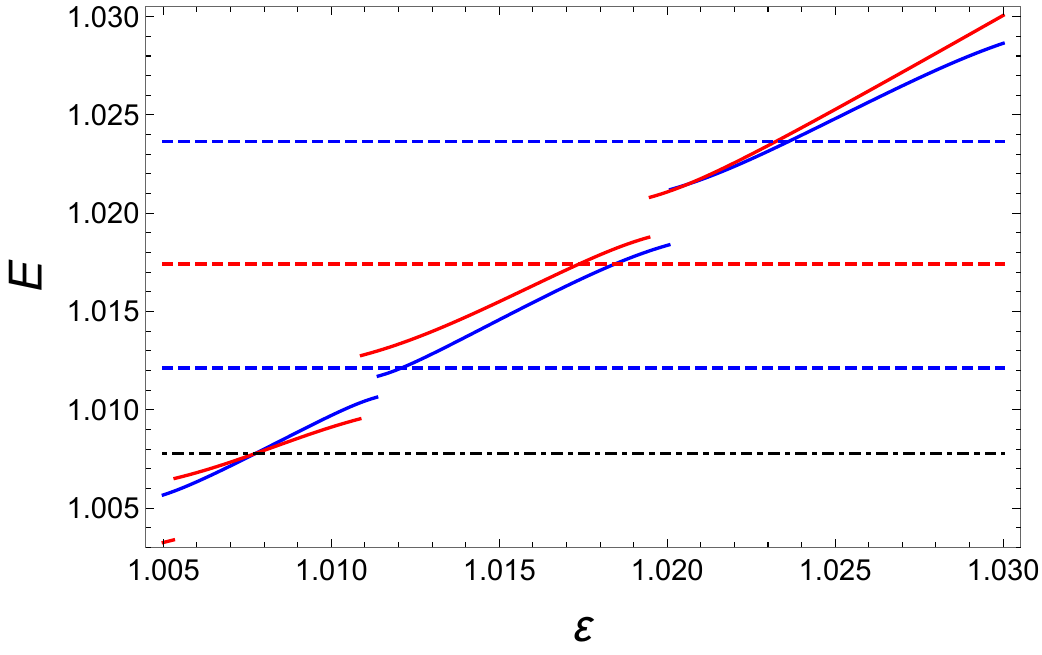}
    \includegraphics[width=0.45\textwidth]{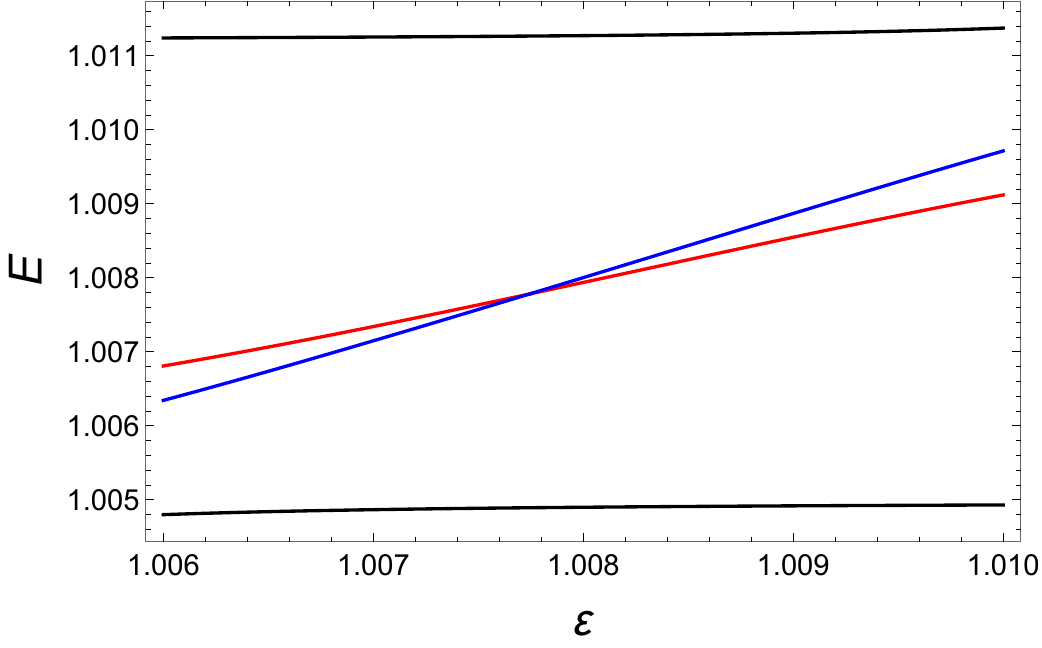}
    \caption{{\textbf{Left panel}.} %MDPI: We have removed the italic formatting and added bold formatting. Please confirm. Same as below
    %AUTHORS: Ok everywhere.
    Energy of the eigenstates with highest $P_e$ in the symmetric (solid blue line) and antisymmetric (solid red line) sector, for~$\gamma=10^{-4}$, $L=40\pi$ and $d=8\pi$. The~horizontal lines, reported for comparison, represent resonant energies in the considered range, corresponding to symmetric (dashed blue lines) and antisymmetric (dashed red line) states, or~to states in both symmetry sectors (dot-dashed black line). In~the last case, a~degeneracy occurs, with~states of opposite symmetry corresponding to the coincident resonant eigenvalues $E_1 = \tilde{E}_4$. {\textbf{Right panel}.} Energy of the quasi-degenerate doublet near the degeneracy point. The~plot highlights the gap between the doublet energies and the closest eigenvalues in the spectrum, above~and below (black lines).}
    \label{fig:energies_8pi}
\end{figure}

The results of this investigation, performed in the range \eqref{eq:range}, are reported in Figure~\ref{fig:energies_8pi}. The~eigenvalues correspond to the highest-$P_e$ eigenstates in the symmetric (blue curves) and antisymmetric (red curves) sectors. A~resonant eigenstate occurs whenever the energy crosses one of the horizontal lines of the same color. An~exception is represented by the dot-dashed line at the double resonance energy $E_1=\tilde{E}_4=1.0078$, where resonant eigenstates can occur in both symmetry sectors. It is evident from the left panel that a degeneracy occurs for a value of the excitation energy that is very close to $E_1$ (and simultaneously $\tilde{E}_4$). The~right panel, viewed within a restricted range around the crossing, shows that the doublet becomes quasi-degenerate, since the closest eigenvalues (black curves above and below) remain well separated for a finite~range.

\begin{figure}[]
%    \centering
    \subfigure[~$\varepsilon=1.0077$, $E=1.0078=E_1$, $P_e=0.8606$]{\includegraphics[width=0.22\textwidth]{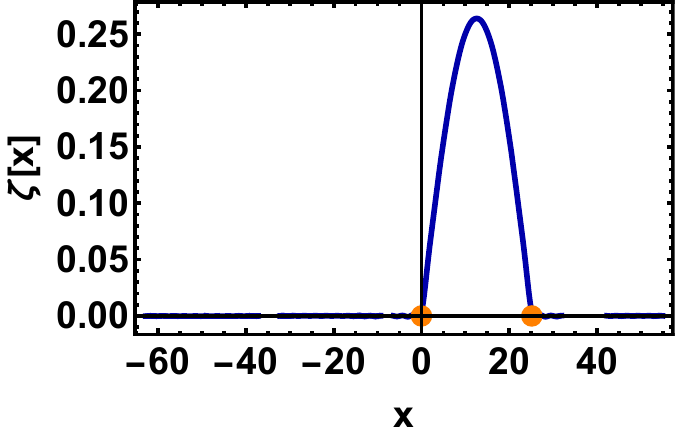}}
    \subfigure[~$\varepsilon=1.0077$, $E=1.0078=\tilde{E}_4$, $P_e=0.6068$]{\includegraphics[width=0.22\textwidth]{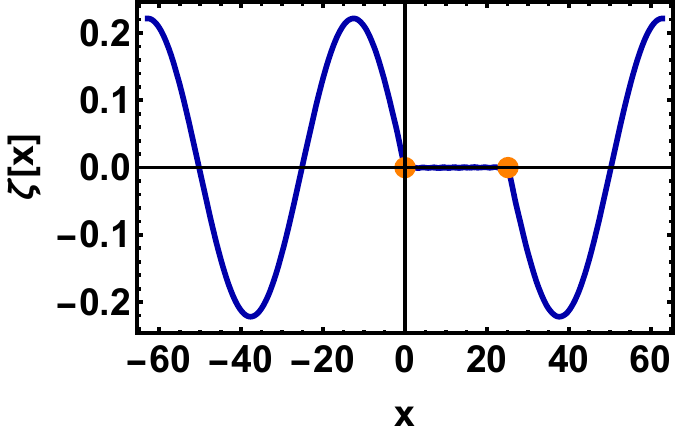}}\\
    \subfigure[~$\varepsilon=1.0121$, $E=1.0121=\tilde{E}_5$, $P_e=0.7091$]{\includegraphics[width=0.22\textwidth]{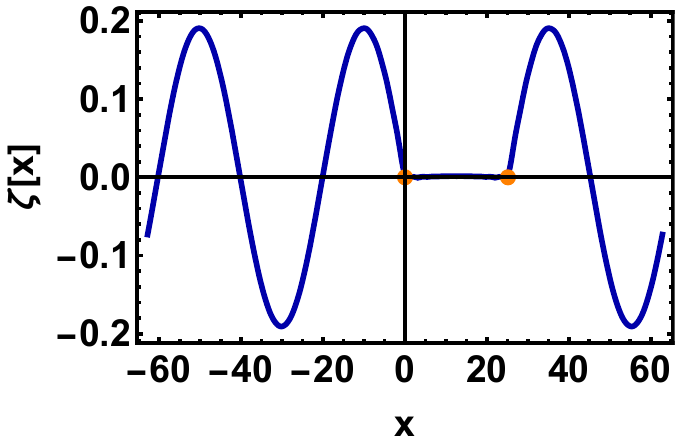}}
    \subfigure[~$\varepsilon=1.0121$, $E=1.0109$, $P_e=0.2163$]{\includegraphics[width=0.22\textwidth]{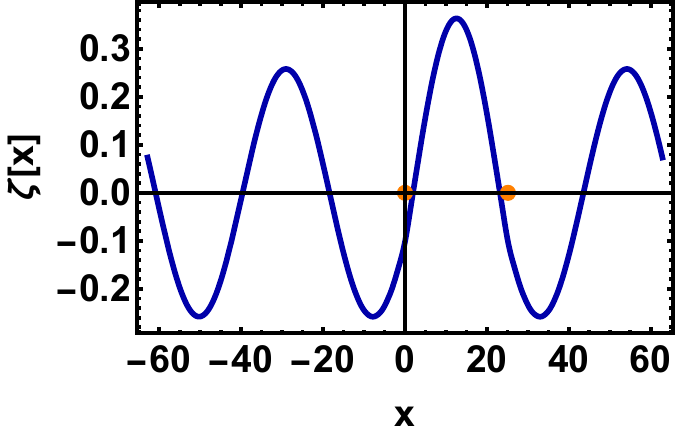}}\\
    \subfigure[~$\varepsilon=1.0174$, $E=1.0174=\tilde{E}_6$, $P_e=0.7733$]{\includegraphics[width=0.22\textwidth]{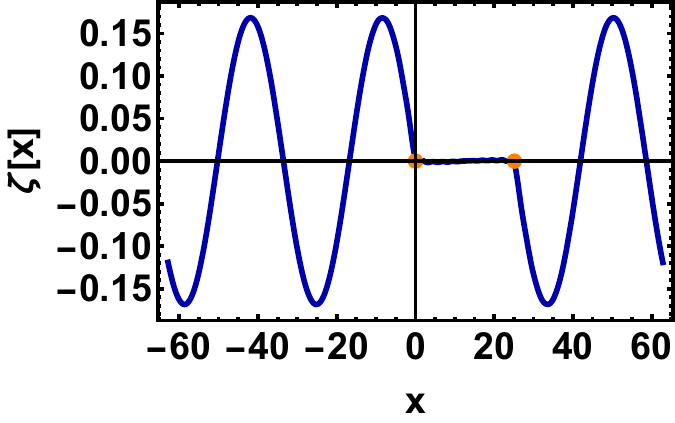}}
    \subfigure[~$\varepsilon=1.0174$, $E=1.0166$, $P_e=0.8006$]{\includegraphics[width=0.22\textwidth]{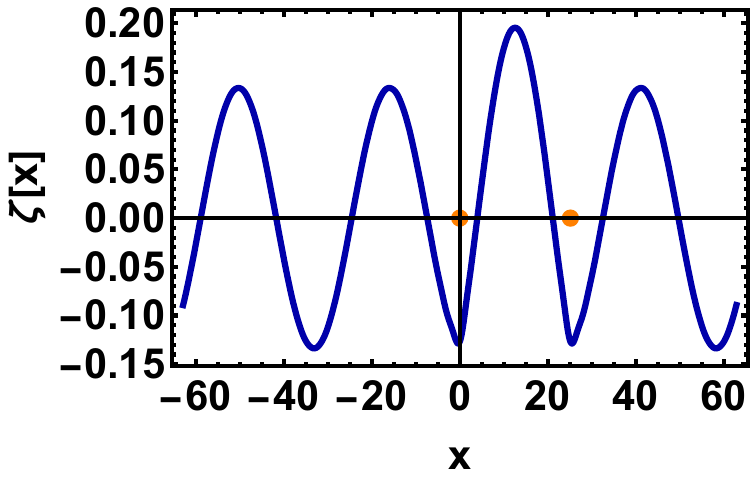}}\\
    \subfigure[~$\varepsilon=1.0236$, $E=1.0237=\tilde{E}_7$, $P_e=0.8230$]{\includegraphics[width=0.22\textwidth]{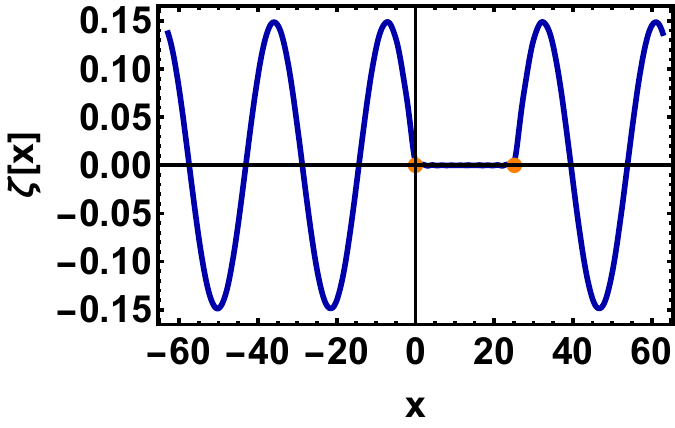}}
    \subfigure[~$\varepsilon=1.0236$, $E=1.0240$, $P_e=0.9130$]{\includegraphics[width=0.22\textwidth]{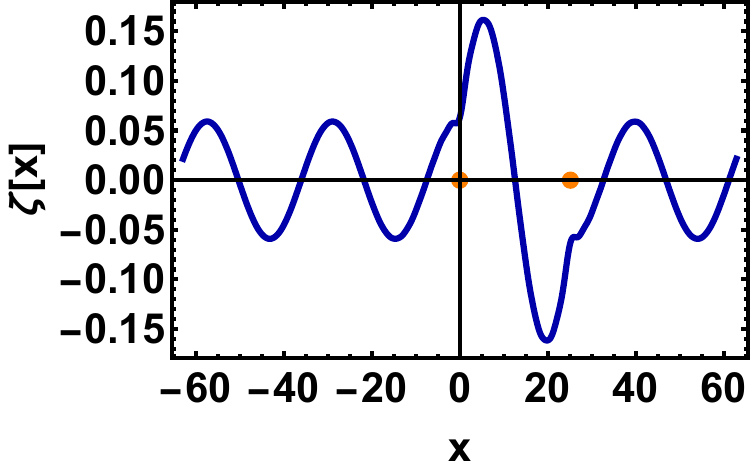}}
    \caption{{Panels (\textbf{a},\textbf{c},\textbf{e},\textbf{g}).} Photon spatial amplitudes of eigenstates that maximize $P_e$ either in the symmetric or the antisymmetric sector, whose eigenvalue $E$ coincide with one of the resonant energies. The~reported plots are referred to $\gamma= 10^{-4}$, $L=40\pi$ and $d=8\pi$. {Panels (\textbf{b},\textbf{d},\textbf{f},\textbf{h}).} Photon spatial amplitudes of the eigenstates with the closest energy to the ones reported on the left. In~the case of panels (\textbf{a},\textbf{b}), the~eigenstates are degenerate, with~$E=E_1=\tilde{E}_4$. The~orange dots highlight the position of the~emitters at $x_1=0$ and $x_2=d=8\pi$.}
    \label{fig:amplitudes_8pi}
\end{figure}

The interpretation of the degenerate eigenstates found for $\varepsilon\simeq E_1$ as a resonant doublet is corroborated by the photon amplitudes represented in Figure~\ref{fig:amplitudes_8pi}. The~cases shown in Figure~\ref{fig:amplitudes_8pi}a,b, %MDPI: We have revised the citation format. Please confirm. The following highlights are the same
%AUTHORS: Ok everywhere. We amended two figure citations in the paragraph below.
which represent the highest-$P_e$ eigenstates in the symmetric and antisymmetric sector, respectively, are similar to those in the upper panels of Figure~\ref{fig:ph-wv1}. Here, however, tuning the value of $\varepsilon$ to achieve degeneracy at $\bar{\varepsilon}\simeq 1.0077$ leads to a full confinement of the photon, either in the shorter [Figure~\ref{fig:amplitudes_8pi}a] or in the longer [Figure~\ref{fig:amplitudes_8pi}b]  path connecting the emitters. Combining this information with that provided by Figure~\ref{fig:energies_8pi}, we conclude that, in~a range $1.006\lesssim\varepsilon\lesssim 1.010$ and at the lowest order in $\delta = \varepsilon - \bar{\varepsilon}$, the~effective dynamics in the resonant doublet sector is determined by the effective Hamiltonian
\begin{equation}
    h_{(1,4)} = \left( \bar{\varepsilon} + c_m \delta \right) \mathrm{I} + c_d \delta \left( |E_1 \rangle \langle E_1| - |\tilde{E}_4\rangle\langle\tilde{E}_4| \right) ,
\end{equation}
where $I$ is the identity on the doublet subspace (generally depending on $\delta$) and $|E_1 \rangle$ and $|\tilde{E}_4\rangle$ represent the states characterized by the spatial amplitudes and $P_e$ reported in Figure~\ref{fig:amplitudes_8pi}a and~\ref{fig:amplitudes_8pi}b, respectively. The~constants $c_{m}$ and $c_d$ correspond to the derivatives with respect to $\delta$ of the mean and the difference, respectively, between~$\langle E_1| H |E_1 \rangle$ and $\langle \tilde{E}_4| H |\tilde{E}_4 \rangle$, accounting for the linear behavior of eigenvalues in Figure~\ref{fig:energies_8pi}. The~remaining panels on the left of Figure~\ref{fig:amplitudes_8pi} show the resonant states found among the highest-$P_e$ ones in their respective sector, with~energy $\tilde{E}_5$ [symmetric, Figure~\ref{fig:amplitudes_8pi}c], $\tilde{E}_6$ [antisymmetric, Figure~\ref{fig:amplitudes_8pi}e] and $\tilde{E}_7$ [symmetric, Figure~\ref{fig:amplitudes_8pi}g]. These states are not paired into doublets: panels on the right report the photon amplitudes of the energetically closest eigenstates, showing that they are neither resonant nor~degenerate.

\unskip

\subsection{Case 2: $d/(L-d)=2/3$}

\begin{figure}[]
%    \centering
    \includegraphics[width=0.45\textwidth]{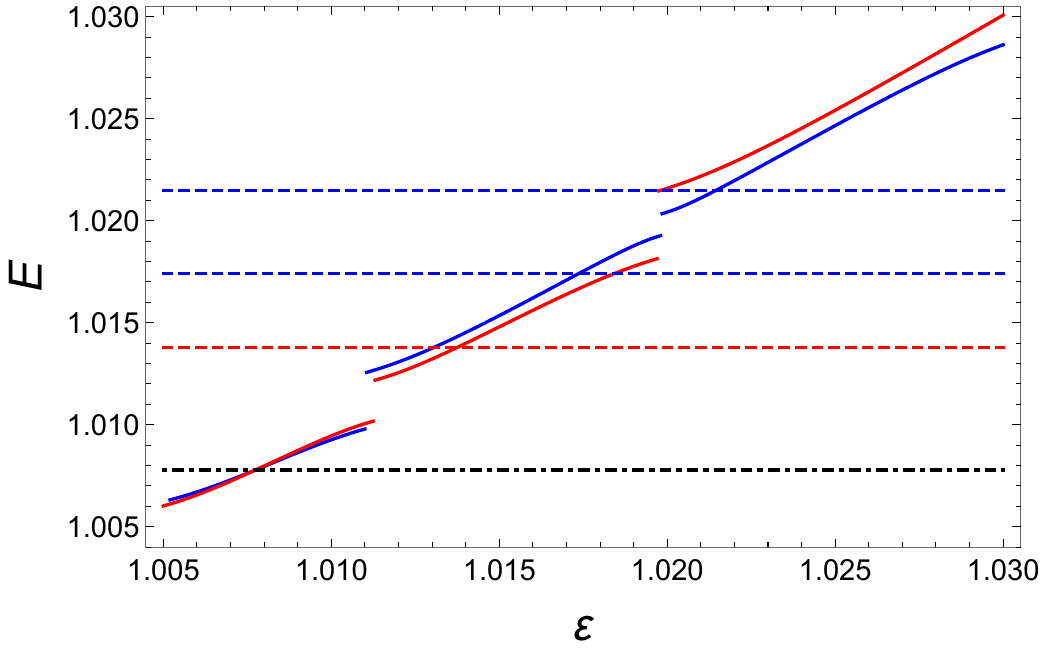}
    \includegraphics[width=0.45\textwidth]{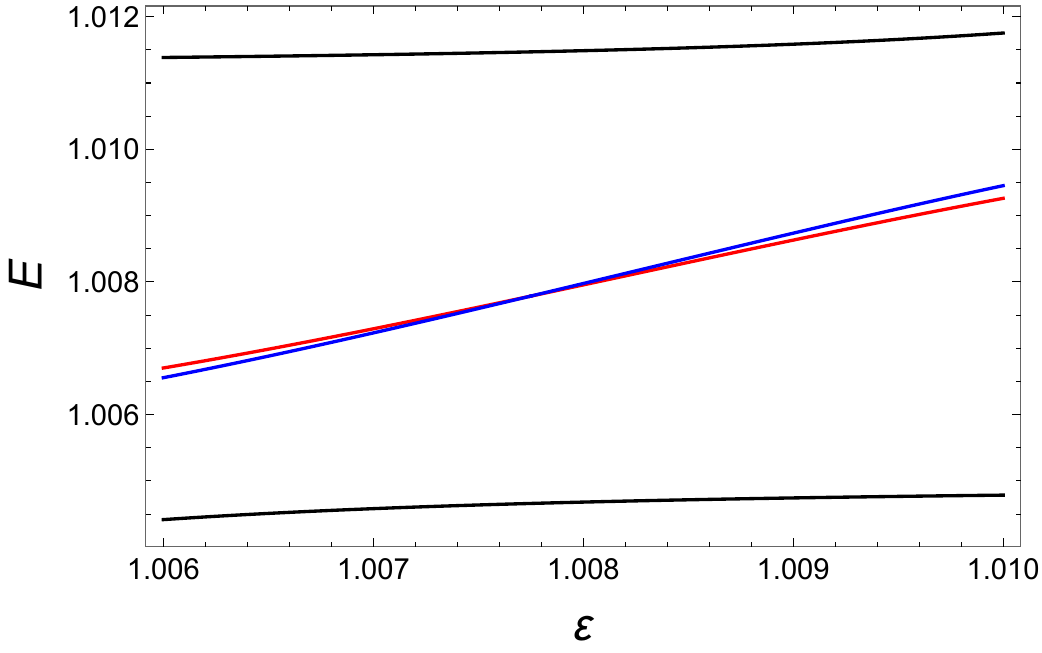}
    \caption{\textbf{Left panel}. Energy of the eigenstates with highest $P_e$ in the symmetric (solid blue line) and antisymmetric (solid red line) sector, for~$\gamma=10^{-4}$, $L=40\pi$ and $d=16\pi$. The~horizontal lines, reported for comparison, represent resonant energies in the considered range, corresponding to symmetric (dashed blue lines) and antisymmetric (dashed red line) states, or~to states in both symmetry sectors (dot-dashed black line). In~the last case, a~degeneracy occurs, with~states of opposite symmetry corresponding to the coincident resonant eigenvalues $E_1 = \tilde{E}_4$. \textbf{Right panel}. Energy of the quasi-degenerate doublet near the degeneracy point. The~plot highlights the gap between the doublet energies and the closest eigenvalues in the spectrum, above~and below (black lines).}
    \label{fig:energies_16pi}
\end{figure}

As a second case study, we keep the waveguide length unchanged, and~consider the~distances
\begin{equation}
    d=16\pi, \quad L-d=24\pi ,
\end{equation}
giving a value of $2/3$ for the ratio of the shorter path to the longer path between the emitters. We repeat the analysis performed in the first case, locating the resonant eigenstates in the same range (\ref{eq:range}) of the emitter excitation energy. The~left panel of Figure~\ref{fig:energies_16pi} reports the energies of the highest-$P_e$ eigenstates in the symmetric (blue curves) and antisymmetric (red curves) sector. In~addition to the double resonance at $E_2=\tilde{E}_3 = 1.0078$, the~eigenvalue curves cross two resonances in the symmetric sector, namely at $E_3=1.0174$ and $\tilde{E}_5=1.0215$, while an antisymmetric resonance is encountered at $\tilde{E}_4 = 1.0138$. The~right panel of Figure~\ref{fig:energies_16pi} offers a magnified view of a neighborhood of the degeneracy point $\bar{\varepsilon}$, along with the two eigenvalues closest to the quasi-degenerate resonant doublet. The~doublet is still locally described by an effective Hamiltonian in the form
\begin{equation}
    h_{(2,3)} = \left( \bar{\varepsilon} + c_m \delta \right) \mathrm{I} + c_d \delta \left( |E_2 \rangle \langle E_2| - |\tilde{E}_3\rangle\langle\tilde{E}_3| \right) , 
\end{equation}
but compared to the case of $d=8\pi$, reported in Figure~\ref{fig:energies_8pi}, it is evident that the coefficient $c_d$ is significantly smaller. This is due to the fact that, as~the $d/(L-d)$ ratio gets closer to one, the~two complementary eigenstates become increasingly similar to each other in their~structure.
%\unskip

\begin{figure}[H]
%    \centering
    \subfigure[~$\varepsilon=1.0077$, $E=1.0078=E_2$, $P_e=0.7553$]{\includegraphics[width=0.22\textwidth]{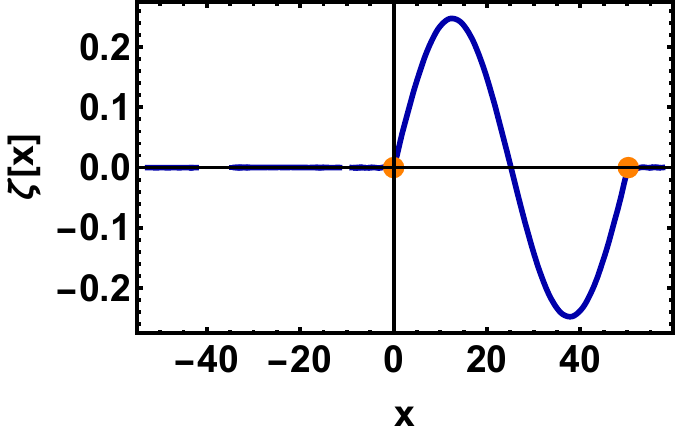}}
    \subfigure[~$\varepsilon=1.0077$, $E=1.0078=\tilde{E}_3$, $P_e=0.6729$]
    {\includegraphics[width=0.22\textwidth]{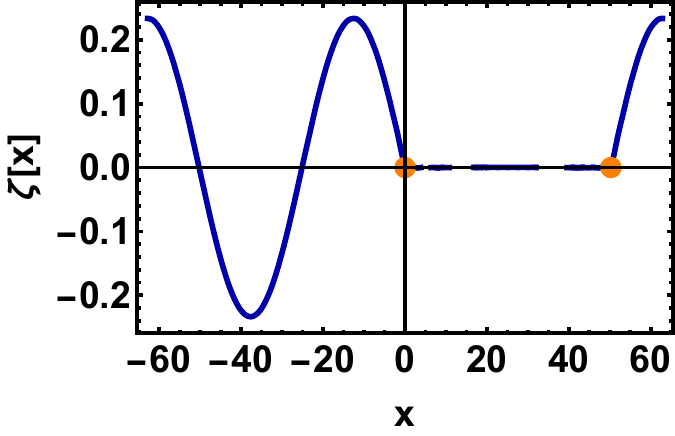}}\\
    \subfigure[~$\varepsilon=1.0138$, $E=1.0138=\tilde{E}_4$, $P_e=0.7868$]{\includegraphics[width=0.22\textwidth]{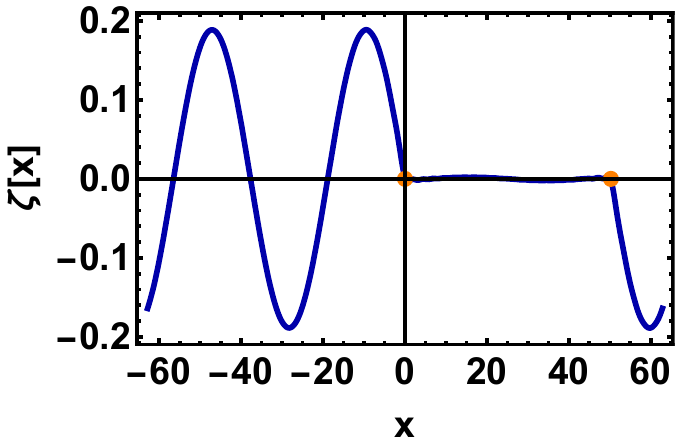}}
    \subfigure[~$\varepsilon=1.0138$, $E=1.0143$, $P_e=0.7930$]{\includegraphics[width=0.22\textwidth]{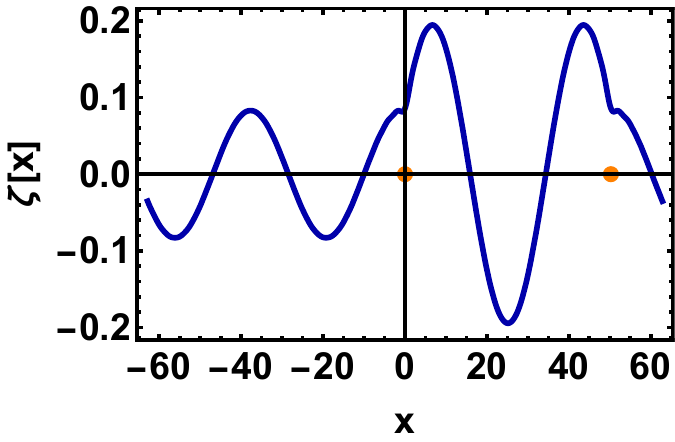}}\\
    \subfigure[~$\varepsilon=1.0174$, $E=1.0174=E_3$, $P_e=0.8726$]{\includegraphics[width=0.22\textwidth]{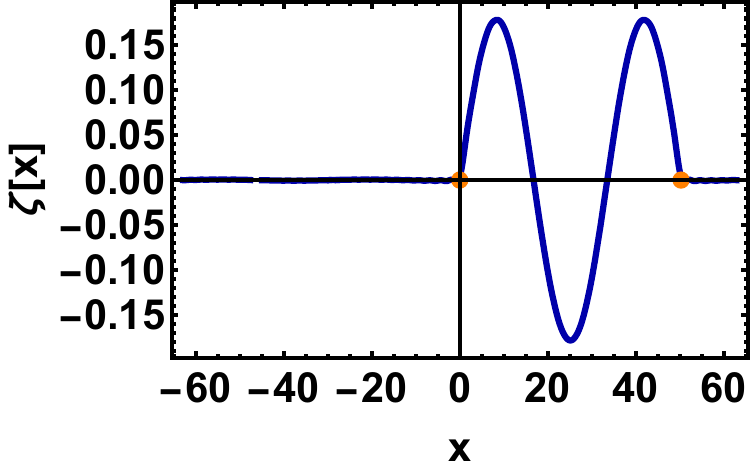}}
    \subfigure[~$\varepsilon=1.0174$, $E=1.0167$, $P_e=0.7385$]{\includegraphics[width=0.22\textwidth]{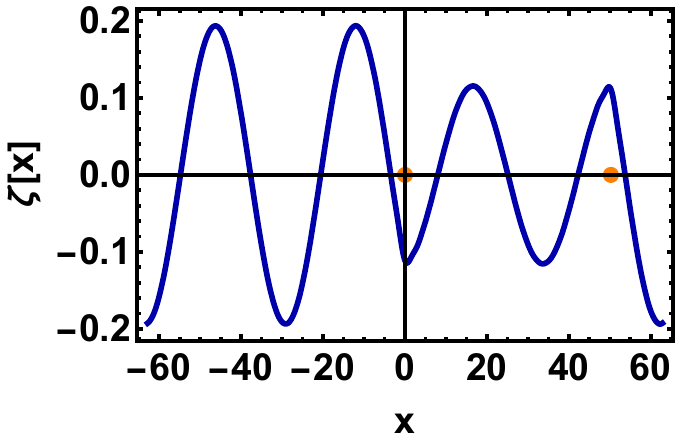}}\\
    \subfigure[~$\varepsilon=1.0214$, $E=1.0215=\tilde{E}_5$, $P_e=0.8474$]
    {\includegraphics[width=0.22\textwidth]{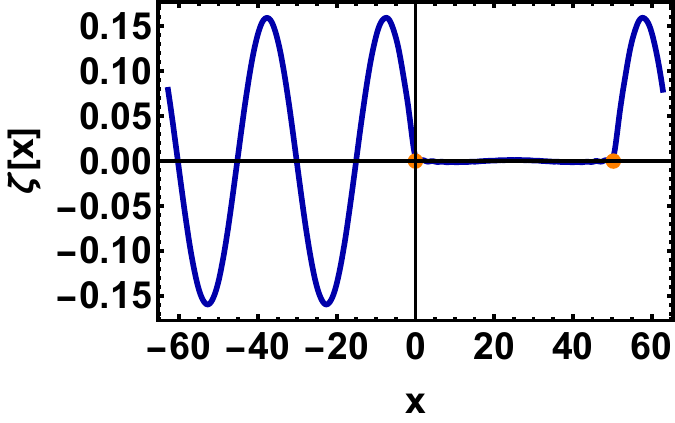}}
    \subfigure[~$\varepsilon=1.0214$, $E=1.0225$, $P_e=0.7001$]{\includegraphics[width=0.22\textwidth]{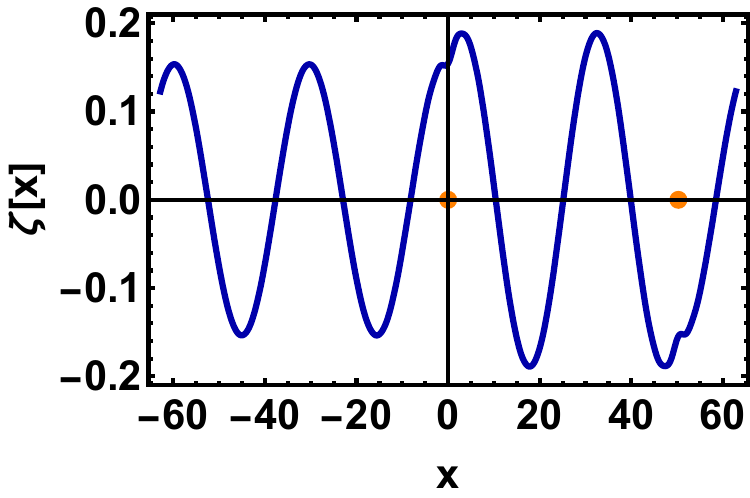}} 
    \caption{{Panels (\textbf{a},\textbf{c},\textbf{e},\textbf{g}).} Photon spatial amplitudes of eigenstates that maximize $P_e$ either in the symmetric or the antisymmetric sector, whose eigenvalue $E$ coincide with one of the resonant energies. The~reported plots are referred to $\gamma= 10^{-4}$, $L=40\pi$ and $d=16\pi$. {Panels (\textbf{b},\textbf{d},\textbf{f},\textbf{h}).} Photon spatial amplitudes of the eigenstates with the closest energy to the ones reported on the left. In~the case of panels (\textbf{a},\textbf{b}), the~eigenstates are degenerate, with~$E=E_2=\tilde{E}_3$. The~orange dots highlight the position of the~emitters at $x_1=0$ and $x_2=d=16\pi$.}
    \label{fig:amplitudes_16pi}
\end{figure}

The photon amplitudes of the two states in the resonant doublet with $\nu=2$ and $\nu'=3$ are represented in Figure~\ref{fig:amplitudes_16pi}a,b. Figure~\ref{fig:amplitudes_16pi}c,e,g show the other resonant states found for $\varepsilon$ in the range \eqref{eq:range}: none of them belongs to a quasi-degenerate resonant doublet, as~one can observe from Figure~\ref{fig:amplitudes_16pi}d,f,h, showing the photon amplitudes of the eigenstates with the closest~energy.

More generally, when the ratio
\begin{equation}
    \frac{d}{L-d} = \frac{\nu}{\nu'}
\end{equation}
is irreducible, we expect a ``fundamental'' doublet, whose internal dynamics is \mbox{described by}
\begin{equation}\label{eq:doubletham}
    h_{(\nu,\nu')} = \left( \bar{\varepsilon} + c_m \delta \right) \mathrm{I} + c_d \delta \left( |E_{\nu} \rangle \langle E_{\nu}| - |\tilde{E}_{\nu'}\rangle\langle\tilde{E}_{\nu'}| \right) 
\end{equation}
when the excitation energy is close to
\begin{equation}
    \bar{\varepsilon} \simeq \sqrt{ \left( \frac{\nu\pi}{d} \right)^2 + m^2 } .
\end{equation}
Any other possible doublet has a larger energy, and~is therefore expected, at~least in a perturbative regime, to~occur at a higher $\varepsilon$.

\section{Conclusions and~Outlook}\label{sec:conclusions}

In both of the considered case studies, we identified a resonant doublet of quasi-degenerate states. Due to the small-coupling regime, the~degeneracy point is found in correspondence of an emitter excitation energy, only slightly displaced with respect to the double resonance energy, making the identification of the energy range where the doublet is expected particularly easy. The~analysis of the spectrum in proximity to the degeneracy point revealed that the gap with the closest eigenvalue remains much larger than the internal energy separation of the doublet for a wide range of excitation energy. This difference is more evident in the case where the ratio between the lengths of the two paths connecting the emitters is closer to 1 (but still strictly smaller than 1).

These features make a quasi-degenerate doublet addressable as a two-level system, in~a suitable coupling energy range, opening the way to possible applications as logical qubits. As~envisaged in Ref.~\cite{Lonigro2021a}, the~states of the doublet, representing the computational basis, could be efficiently discriminated by detecting the photon in either of the two paths, therefore without directly interacting with the emitters. External pulses on the emitters can instead be used to coherently drive one state of the doublet to the other, creating an effective coupling of the ``$\sigma_x$'' and ``$\sigma_y$'' type, which adds to the ``$\sigma_z$'' term already present in the doublet Hamiltonian \eqref{eq:doubletham}. It is worth remarking that, while the results in this article are obtained under the assumption of spatial symmetries, breaking those symmetries can provide a tool to couple the states inside a doublet and implement the whole set of observables in a two-dimensional space. If~the strength of symmetry-breaking terms is low, they are not expected to qualitatively alter the doublet picture in Figures~\ref{fig:energies_8pi}--\ref{fig:energies_16pi}, besides~generally transforming the level crossing at the degeneracy point into an avoided~crossing.

The implementation of such a procedure will be the subject of future research. The~reference platform to realize the proposed scheme is represented by circuit quantum electrodynamics: on one hand, the~typical wavelengths involved in the dynamics of superconducting emitters coupled through resonators and transmission lines are in the microwave range, ensuring the feasibility of the length matching conditions; on the other hand, the~scheme provides the required flexibility to both realize closed geometries, and~implement outcoupling and interactions with classical fields~\cite{hoi2015probing,nakamura2020programmable,kannan2020generating}. While resonant doublets can hardly be considered a universal solution for a qubit implementation due to scalability issues, they still can provide a stable platform of interest for few-qubit protocols. Therefore, in~a wider perspective, it would be interesting to investigate the possibility to coherently manipulate and entangle systems of two or more closed waveguides, hosting a resonant doublet at the same energy, by~connecting them through microwave transmission~lines.

\section*{Acknowledgements}

The authors thank Domenico Pomarico for useful~discussions. This research was funded by Ministero dell'Università e della Ricerca (MUR) PNRR projects CN00000013 ``National Centre on HPC, Big Data and Quantum Computing'' and PE0000023 ``National Quantum Science and Technology Institute (NQSTI)''.

\bibliography{references}

\end{document}